\documentclass[aps,preprint,floatfix,nofootinbib,showpacs]{revtex4}
\usepackage{graphicx,color}
\begin{document}
\renewcommand{\thefootnote}{\fnsymbol{footnote}}

\title{Light Pseudoscalar Higgs boson in Neutralino Decays in the
Next-to-Minimal Supersymmetric Standard Model}

\author{
Kingman Cheung$^{1,2}$ and Tie-Jiun Hou$^1$}
\affiliation{
$^1$ Department of Physics, National Tsing Hua University,
Hsinchu, Taiwan \\
$^2$
Physics Division, National Center for Theoretical Sciences, Hsinchu, Taiwan
}
\date{\today}

\begin{abstract}
We point out another important production channel of a light
pseudoscalar Higgs boson $a_1$ via the decays of neutralinos,
including $\widetilde{\chi}_{2,3}^0 \to \widetilde{\chi}_{1}^0 a_1$,
in the framework of the NMSSM.  We scan the whole parameter space
using the most up-to-date version of NMHDECAY and search for regions
where $B(\widetilde{\chi}_{2,3}^0 \to \widetilde{\chi}_{1}^0 a_1) >
0.5$.  If the gluino and squarks are light enough for copious
production of SUSY events at the LHC, there would be numerous number
of $\widetilde{\chi}_{2,3}^0$ in subsequent decays of gluinos and
squarks.  Therefore, the production rates of $a_1$ via neutralino
decays would be more important than $h \to a_1 a_1$ and others.
Potentially, the final state is filled with many $\tau$ leptons, 
which can be reconstructed at the mass of $a_1$.  
This is a clean, observable, and 
distinguishable signature for NMSSM.
\end{abstract}
\maketitle

\section{Introduction}
Supersymmetry is the leading candidate for the physics beyond the
standard model (SM).  It not only solves the gauge hierarchy problem,
but also provides a dynamical mechanism for electroweak symmetry
breaking and a natural candidate for the dark matter.  The minimal
version, the minimal supersymmetric standard model (MSSM), has
suffered from what so called little hierarchy problem and the $\mu$
problem.

An extension with an extra singlet superfield, known as the
next-to-minimal supersymmetric standard model (NMSSM) was motivated to
provide a natural solution to the $\mu$ problem.  The $\mu$ parameter
in the term $\mu H_u H_d$ of the superpotential of the MSSM naturally
has its value at either $M_{\rm Planck}$ or zero (due to a symmetry).
However, the radiative electroweak symmetry breaking conditions
require the $\mu$ parameter to be of the same order as $m_Z$ for
fine-tuning reasons. Such a conflict was coined as the $\mu$ problem
\cite{mu-problem}.  In the NMSSM, the $\mu$ term is generated dynamically
through the vacuum-expectation-value (VEV), $v_s$, of the scalar
component of the additional Higgs field $S$, which is naturally of the
order of the SUSY breaking scale.  Thus, an effective $\mu$ parameter
of the order of the electroweak scale is generated.

The NMSSM was recently revived because it was shown that it can
effectively relieve the little hierarchy problem \cite{derm}.  Due to
the additional Higgs singlet field and an approximate PQ symmetry, the
NMSSM naturally has a light pseudoscalar Higgs boson $a_1$.  It has been
shown \cite{derm} that, in most parameter space that is natural, the
SM-like Higgs boson can decay into a pair of light pseudoscalar bosons
with a branching ratio larger than $0.7$.  Thus, the branching ratio
of the SM-like Higgs boson into $b\bar b$ would be less than $0.3$ and
so the LEPII bound is effectively reduced to around 100 GeV
\cite{LEP}. 

The dominance of $h\to a_1 a_1$ mode for the intermediate Higgs boson
has significant impacts on the Higgs search strategies \cite{ellw}.
The most useful channel for intermediate Higgs boson, $h\to
\gamma\gamma$ and $h \to W W^*$ would be substantially affected
because $B(h\to\gamma\gamma)$ lowers by a factor of a few.  So is the
$h\to b \bar b$ in $Wh,Zh$ production.  New search modes via $h \to
a_1 a_1$ are mandatory.  For example, $h \to a_1 a_1 \to 4b$ for
$m_{a_1} > 2 m_b$ via $Wh, Zh$ production with at least one charged
lepton and $4 B$-tags in the final state has been shown to afford a
clean signal of high significance and a full Higgs mass reconstruction
at the LHC \cite{qs}. The associated Higgs production with gauge bosons
or $t\bar t$ pairs was shown to be effective \cite{moretti}.
Similar studies at the Tevatron were also
performed \cite{scott}.  Other possibilities like $h \to a_1 a_1 \to 2b 2\tau$
\cite{carena} and $h \to a_1 a_1 \to 4\tau$ \cite{4tau,carena} can
further enhance the signal, especially when $ 2 m_\tau < m_{a_1} <
2m_b$.  In the extreme limit of zero mixing with the MSSM
pseudoscalar, the singlet-like $a_1$ can decay into a pair of
photons. In this case, $h \to a_1 a_1 \to 4 \gamma$ \cite{matchev}.
The light pseudoscalar $a_1$ can also be produced in non-Higgs decays.
It can be produced in $B$ meson decays \cite{hiller,Domingo,Heng}, in
$\Upsilon$ decays \cite{bob} and other rare decays \cite{he}, and
in associated production with chargino
pair \cite{assoc}.  In some other contexts, a light pseudoscalar boson
can also be frequently produced in association with a Higgs boson or
a heavy quark pair \cite{song-yan}.  
There could also be other unconventional decay modes of the SM-like
Higgs boson in the NMSSM, e.g., invisible decay into neutralinos \cite{vernon}.
A recent summary can be found in Ref. \cite{Chang:2008cw}.

In this note we point out another important production channel of a
light pseudoscalar Higgs boson $a_1$ via the decays of neutralinos,
including $\widetilde{\chi}_{2,3}^0 \to \widetilde{\chi}_{1}^0 a_1$,
in the framework of the NMSSM.  This is potentially much more
important than from the decay of the Higgs boson or from the
associated production.  In particular, if the gluino and squarks are
light enough for copious production of SUSY events at the LHC, there
would be numerous number of $\widetilde{\chi}_{2,3}^0$ in subsequent
decays of gluinos and squarks.  Therefore, the production rates of
$a_1$ via neutralino decays would be much more important than $h \to
a_1 a_1$ and others.
There is also the possibility of $\widetilde{\chi}_{2,3}^0 \to
\widetilde{\chi}_{1}^0 h$ followed by $h \to a_1 a_1$.  It was argued
in Refs. \cite{derm} that the mass $m_{a_1}$ in most favorable
parameter space is lighter than $2m_b$, and thus $a_1 \to \tau^+
\tau^-$ is the most frequent.  In this case, SUSY events would be
filled with many $\tau$ leptons plus missing energies, with the
corresponding $\tau^+ \tau^-$ reconstructed at the $a_1$ mass.

We scan the whole parameter space using the most up-to-date version of
NMHDECAY \cite{nmhdecay} and search for regions where 
$B(\widetilde{\chi}_{2,3}^0 \to  \widetilde{\chi}_{1}^0  a_1) > 0.5$.
We show characteristics of this region of parameter space.

\section{Two body and three body decays of neutralino}

The superpotential of the NMSSM is given by
\begin{equation}
W = \mathbf{h_u} \hat{Q} \, \hat{H}_u \,\hat{U}^c
- \mathbf{h_d} \hat{Q}  \, \hat{H}_d \, \hat{D}^c
- \mathbf{h_e} \hat{L}  \, \hat{H}_d \, \hat{E}^c
+\lambda \hat{S} \, \hat{H}_u \, \hat{H}_d+ \frac{1}{3}\kappa \, \hat{S}^3.
\end{equation}
where
$\hat{Q}$, $\hat{L}$, $\hat{H}_u$, $\hat{H}_d$, $\hat{U}^c$, $\hat{D}^c$,
$\hat{E}^c$, and $\hat{S}$ 
are the doublet quark and lepton, doublet up-type Higgs and down-type
Higgs, singlet up-quark and down-quark, and the singlet scalar
superfields, respectively.

The Higgs sector of the NMSSM consists of the usual two Higgs doublets 
$H_u$ and $H_d$ and an extra Higgs singlet $S$.
The extra singlet field is allowed to couple
only to the Higgs doublets of the model, the supersymmetrization of which
is that the singlet field only couples to the Higgsino doublets.
Consequently, the couplings of the singlet $S$ to gauge bosons and fermions
will only be manifest via their mixing with the doublet Higgs fields. 
After the Higgs fields take on the VEV's and rotating away the Goldstone
modes, we are left with a pair of charged Higgs bosons, 3 real scalar fields,
and 2 pseudoscalar fields.  In particular, the mass matrix for
 the two pseudoscalar Higgs bosons $P_1$ and $P_2$ is 
\begin{equation}
V_{\rm pseudo} = \frac{1}{2} \;( P_1 \;\; P_2)  {\cal M}^2_P \; 
           \left( \begin{array}{c}
                             P_1 \\
                             P_2 \end{array} \right )
\end{equation}
with
\begin{eqnarray}
{\cal M}^2_{P\, 11} &=& M^2_A \;, \nonumber \\
{\cal M}^2_{P \, 12} &=& {\cal M}^2_{P \, 21} = \frac{1}{2} \cot\beta_s\,
\left(M^2_A \sin 2\beta - 3 \lambda \kappa v_s^2 \right ) \;, \nonumber \\
{\cal M}^2_{P \, 22} &=& \frac{1}{4} \sin 2\beta \cot^2 \beta_s
\,\left( M^2_A \sin 2\beta + 3 \lambda \kappa v_s^2 \right ) \nonumber \\
&& - \frac{3}{\sqrt{2}} \kappa A_\kappa v_s \;, 
\label{pmass}
\end{eqnarray}
where 
\begin{equation}
M_A^2 = \frac{\lambda v_s}{\sin 2\beta}\left(
             \sqrt{2} A_\lambda + \kappa v_s  \right ) \;,
\label{mA}
\end{equation}
and $\tan \beta = v_u/v_d$ and $\tan \beta_s = v_s/v$ and $v^2=v_u^2+v_d^2$.
Here $P_1$ is the pseudoscalar
in MSSM while $P_2$ comes from the singlet $S$ and 
from the effects of rotating away the Goldstone modes. 
The pseudoscalar fields are further rotated to the diagonal basis
($A_1$, $A_2$) through a mixing angle
\cite{miller}:
\begin{eqnarray}
\left( \begin{array}{c} A_2 \\ A_1 \end{array} \right)=
\left( \begin{array}{cc}
     \cos \theta_A & \sin \theta_A \\
     -\sin \theta_A & \cos \theta_A  \end{array} \right)
\left( \begin{array}{c} P_1 \\ P_2 \end{array} \right)
\end{eqnarray}
where the masses of $A_i$ are arranged such that $m_{A_1} < m_{A_2}$.
At tree-level the mixing angle is given by
\begin{equation}
\tan \theta_A = \frac{ {\cal M}^2_{P\,12} }{ {\cal M}^2_{P\,11} - m^2_{A_1}}
 = \frac{1}{2} \cot\beta_s \,
  \frac{M^2_A \sin 2 \beta - 3 \lambda \kappa v_s^2}
  { M^2_A - m^2_{A_1} } \;.
\label{tanA}
\end{equation}
We also use $a_1$ to denote the $A_1$.  The $a_1$ is mainly the singlet
when $\theta_A$ is small.  The couplings of $a_1$ to fermions scale with
$\sin\theta_A$ while $a_1$ can have large couplings to Higgsinos and 
Higgs bosons via the $\lambda S H_u H_d$ term of the superpotential.

\begin{figure}[t!]
\centering
\includegraphics[width=3.2in]{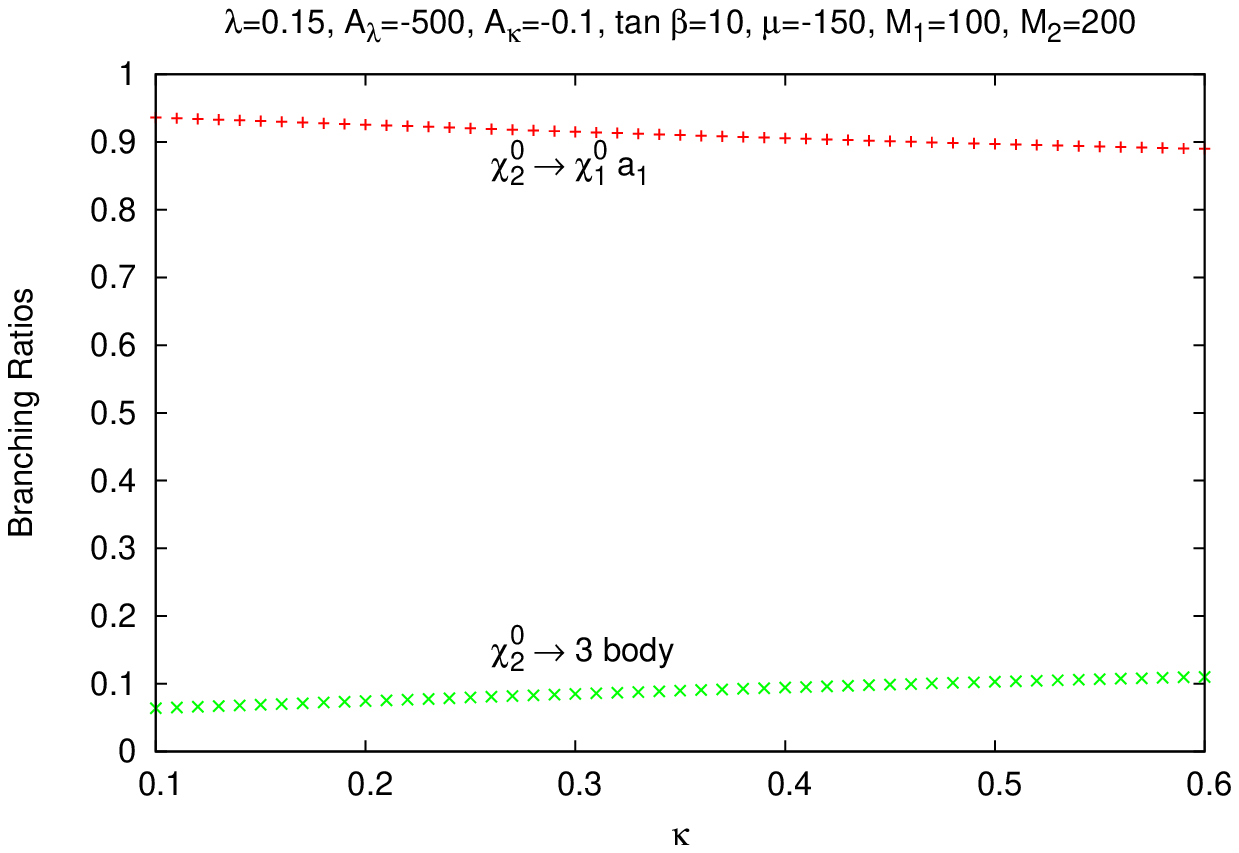}
\includegraphics[width=3.2in]{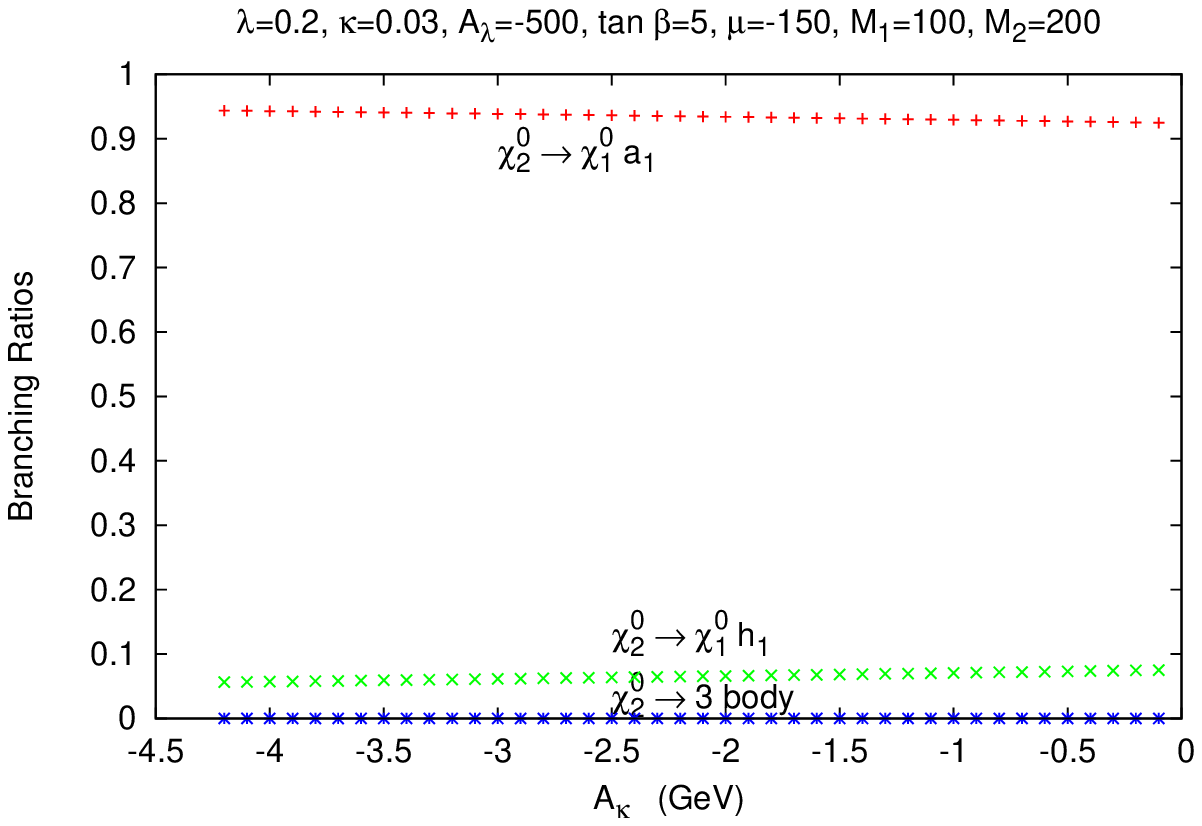}
\caption{\small \label{br1}
Decay branching ratios for 
$\widetilde{\chi}^0_{2} \to \widetilde{\chi}^0_{1} + (a_1, h_1, Z, 3-body)$
with 
(a) $\lambda=0.15$, $A_\lambda=-500$ GeV, $A_\kappa=-0.1$ GeV, $\tan\beta=10$,
$\mu=-150$ GeV, $M_1 =100$ GeV, and $M_2 = 200$ GeV versus $\kappa$;
(b) $\lambda=0.2$, $A_\lambda=-500$ GeV, $\kappa=0.03$, $\tan\beta=5$,
$\mu=-150$ GeV, $M_1 =100$ GeV, and $M_2 = 200$ GeV versus $A_\kappa$.
}
\end{figure}

We use the publicly available code, NMHDECAY \cite{nmhdecay}, to 
generate parameter space points.  Currently, the code has imposed 
a number of experimental constraints, including the radiative
$b\to s\gamma$ decay, the $B_d$ and $B_s$ mixing parameters, 
$B_s \to \mu^+ \mu^-$
decay, $B^+ \to \tau^+ \nu_\tau$ decay, and the relic density of the
lightest neutralino.  These experimental constraints can be turned on or off.
The parameter space points presented in this section satisfy all the above
constraints.  Before we show the decay branching ratios of the second lightest
neutralino, we would like to give the vertex factor that we are considering.
The vertex factor $a_1$-$\widetilde{\chi}^0_i$-$\widetilde{\chi}^0_j$ is 
given by 
\begin{eqnarray}
{\cal L} &=& a_1 \overline{\widetilde{\chi}_i^0} \, i\gamma^5 \,
                 \widetilde{\chi}_j^0 \, \biggr \{
  \frac{g}{2} \, \sin\theta_A\, (N_{j2} - N_{j1} \tan\theta_w ) \,
                     (N_{i4} c_\beta - N_{i3} s_\beta ) \nonumber \\
&& - \frac{\lambda}{\sqrt{2}}\, \sin\theta_A\, ( N_{j3} c_\beta +N_{j4} s_\beta)
\, N_{i5} 
+\frac{\lambda N_{j3} N_{i4} - \kappa N_{i5} \, N_{j5} }{\sqrt{2}} \, 
  \cos\theta_A \,   + 
  (i \leftrightarrow j) \biggr \} \;.
\end{eqnarray}
The other competing channels for 
$\widetilde{\chi}^0_2 \to \widetilde{\chi}^0_1 + X$ include $X=h_1,Z$,
and 3-body decays via off-shell particles.
The detailed formulas will be shown in a future publication \cite{new}.
The 3-body decays are suppressed as long as at least one of the 2-body
modes are open.  We show in Fig. \ref{br1} the branching ratios of
$\widetilde{\chi}^0_{2} \to \widetilde{\chi}^0_{1} + (a_1, h_1, Z, 
3-body)$ for two
sets of parameter space points, with varying $\kappa$ and $A_\kappa$ 
respectively. It is clear that
$B(\widetilde{\chi}^0_{2} \to \widetilde{\chi}^0_{1} + a_1)$ dominates
in these two sets of points,
which satisfy all experimental constraints and relic density of the LSP.
The parameters $\kappa$ and $A_\kappa$ can be kept small by the approximate
PQ and $R$ symmetries, which guarantee the lightness of $a_1$. 

Experimentally, the decay 
$\widetilde{\chi}^0_{2} \to \widetilde{\chi}^0_{1} + a_1$
 gives very interesting signatures.  In typical SUSY
events, a lot of gluinos or squarks are produced, which subsequently decay
into the third or second lightest neutralinos, 
instead of directly decaying into
the lightest one.  Thus, there are numerous second lightest or
third lightest neutralinos, which then decays into the lightest 
neutralino and the light pseudoscalar $a_1$.  The $a_1$ then decays into 
either $b\bar b$ or $\tau^+ \tau^-$ depending on its mass.  Therefore, the
final state will be filled with many $\tau$ leptons, which can be carefully
reconstructed at the mass of $a_1$.  This is a clean, observable, and 
distinguishable signature for NMSSM.
\footnote
{In gauge-mediated models, excessive production of $\tau$ leptons
is often the signature.  However, it can be distinguished between the 
gauge-mediated models and the NMSSM, because the right $\tau$ pair in NMSSM
can be reconstructed at $m_{a_1}$ but not for the $\tau$ leptons 
in GMSB models.}

\section{Decays of squark into neutralinos}

\begin{figure}[t!]
\centering
\includegraphics[width=5.5in]{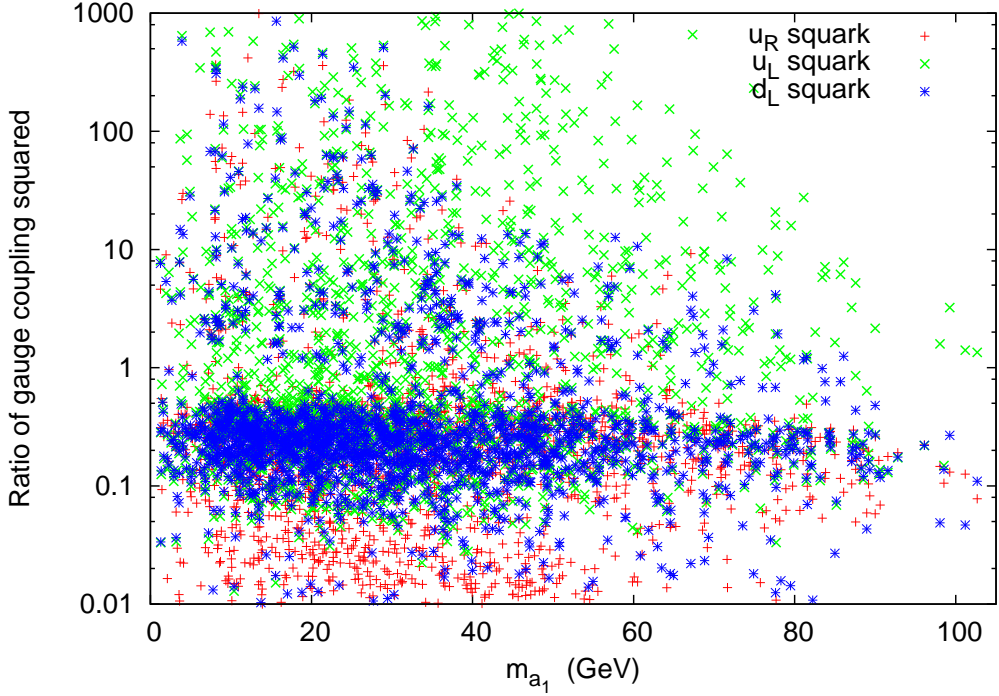}
\caption{\small \label{squark-decay}
Relative branching ratios of the left-handed or right-handed squark into
a quark and the neutralinos:
$|g_{\widetilde{q}_{L,R}\,q_{L,R} \,\widetilde{\chi}^0_{2}}/
g_{\widetilde{q}_{L,R} \, q_{L,R} \,\widetilde{\chi}^0_{1}}|^2$.
Under the requirements of NMHDECAY and 
$B(\widetilde{\chi}^0_2 \to \widetilde{\chi}^0_{1} a_1 ) > 0.5$.
}
\end{figure}

Here we show the relative branching ratios of a squark decaying into
the lightest and second lightest neutralinos
\begin{equation}
\widetilde{q}_{L,R} \to q_{L,R} \; \widetilde{\chi}^0_{1,2}
\end{equation}
where we simply assume that the Yukawa coupling of the light quark is
negligible compared with the gauge couplings.  
The couplings of
$\widetilde{q}_{L,R}$-$q_{L,R}$-$\widetilde{\chi}^0_{i}$ are given by
\begin{eqnarray}
g_{\widetilde{u}_{L}u_{L}\widetilde{\chi}^0_{i}} =
 -\frac{g}{\sqrt{2}} \left( N_{i2} + \frac{t_w}{3} N_{i1} \right )\;, & \qquad &
g_{\widetilde{u}_{R}u_{R}\widetilde{\chi}^0_{i}} = 
                      \frac{g t_w}{\sqrt{2}} \frac{4}{3} N_{i1}\;,  \nonumber \\
g_{\widetilde{d}_{L}d_{L}\widetilde{\chi}^0_{i}} =
 -\frac{g}{\sqrt{2}} \left(- N_{i2} + \frac{t_w}{3} N_{i1} \right)\;, & \qquad &
g_{\widetilde{d}_{R}d_{R}\widetilde{\chi}^0_{i}} = 
                     - \frac{g t_w}{\sqrt{2}} \frac{2}{3} N_{i1} \;.
\end{eqnarray}
We scan the parameter space points with the requirements defined by NMHDECAY, 
except for the relic density of the lightest neutralino, 
and $B(\widetilde{\chi}^0_2 \to \widetilde{\chi}^0_1 a_1) > 0.5$ (details in 
the next section).  
With these parameter space points we calculate the squares of the 
relative gauge couplings:
\[
|g_{\widetilde{q}_{L,R}\,q_{L,R}\widetilde{\chi}^0_{2}}/
g_{\widetilde{q}_{L,R}\,q_{L,R}\widetilde{\chi}^0_{1}}|^2
\]
which can roughly indicate the relative branching ratios without taking
into account the masses in the final state.
We show in Fig. \ref{squark-decay} the squares of relative
gauge couplings.
It is clear that the squarks, whether left-handed or right-handed,
can frequently decay into the second lightest neutralino,
instead of just directly into the lightest one.
Therefore, it is consistent with what we have pointed out in the Introduction
that there are numerous second or even the third lightest neutralinos via
production of squarks or gluinos in SUSY events.  They will then decay into
the lightest neutralino and the light pseudoscalar Higgs boson $a_1$.
We will map the regions in parameter space that the branching 
ratio $B(\widetilde{\chi}^0_{2,3} \to \widetilde{\chi}^0_1 \, a_1) > 0.5$
in the next section.

\section{Scan of NMSSM parameter space}
\begin{figure}[t!]
\centering
\includegraphics[width=5in]{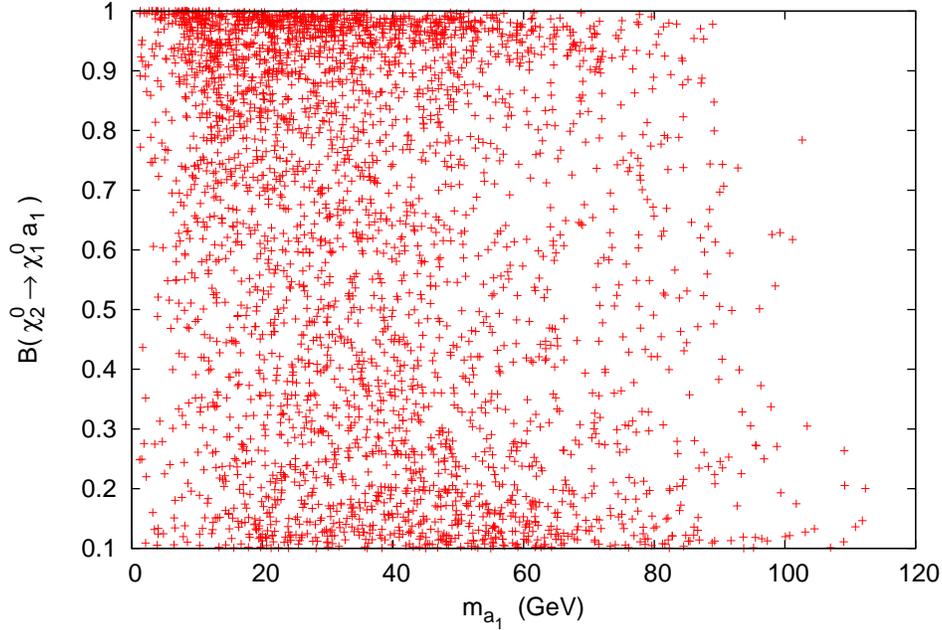}
\caption{\small \label{scatter}
Scatter plot in the plane of $m_{a_1}$ and  $B(\widetilde{\chi}^0_2
\to \widetilde{\chi}^0_1 \,a_1)$.  The points pass the criteria of NMHDECAY
and the branching ratio larger than $0.1$.
}
\end{figure}

Here we scan for the parameter space using the most up-to-date version
of NMHDECAY \cite{nmhdecay} in the following ranges of parameters
\begin{eqnarray}
\lambda:\; 0\; - \;0.7\,, && A_\lambda:\; -1000\; -\; 1000\;{\rm GeV}
  \,,\nonumber\\
\kappa:\;  -0.7 \;-\; 0.7\,, && A_\kappa:\; -10\; -\; 10 \;{\rm GeV}
   \,,\nonumber \\
\tan\beta: \; 1 \;-\; 40  \,, && \mu:\; -500\; -\; 500\;{\rm GeV} 
\,,\nonumber \\
 M_1:\; 0 \;-\;1000 \; {\rm GeV}\,, && M_2:\; 0 \;-\; 1000\; {\rm GeV}\,, 
\end{eqnarray}
where $M_1$ and $M_2$ are the bino and wino mass parameter, respectively.
We ran for a total of 10 million random points in the parameter space. 
The successful points have to pass the criteria of NMHDECAY, including
LEPII bounds, $b\to s\gamma$, $B_d$ and $B_s$ mixing, $B_s \to \mu^+ \mu^-$
and $B^+ \to \tau^+ \nu_\tau$, but not the LSP relic density.  From
the pool of successful points we then calculate the branching ratio of
$B(\widetilde{\chi}^0_{2} \to  \widetilde{\chi}^0_{1} \, a_1)$ and pick those
with the branching ratio larger than $0.1$. 
These points are shown in Fig.~\ref{scatter}.  It is easy to see those 
points with the branching ratio larger than $0.5$ in the figure.
We also show in Table \ref{cuts} the reduction of the number of points
under the requirements of NMHDECAY, and further under 
$B(\widetilde{\chi}^0_{2} \to \widetilde{\chi}^0_{1} + a_1) > 0.1$ and $0.5$,
respectively.

\begin{table}[b!]
\caption{\label{cuts}
Total number of points used, that after scanned by NMHDECAY, and that
after imposing $B( \widetilde{\chi}^0_2 \to \widetilde{\chi}^0_1 a_1) 
> 0.1$ and $0.5$.
}
\begin{tabular}{lr}
\hline
 Steps &  Number of points \\
\hline
Total used  & 10,000,000 \\
Scanned after NMHDECAY & 41,318 \\
$B( \widetilde{\chi}^0_2 \to \widetilde{\chi}^0_1 a_1) >0.1$ & 3,260 \\
$B( \widetilde{\chi}^0_2 \to \widetilde{\chi}^0_1 a_1) >0.5$ & 2,030 \\
\hline
\end{tabular}
\end{table}

\section{Conclusions}
In this note we have scanned the parameter space of the NMSSM with the 
help of the code NMHDECAY.  In a sizable fraction of the parameter space
points, shown in Table \ref{cuts}, the branching ratio of
$B(\widetilde{\chi}^0_{2} \to \widetilde{\chi}^0_{1} + a_1) > 0.5$.
It indicates a potentially more important production channel of the
light pseudoscalar Higgs boson $a_1$ via the decays of neutralinos,
including $\widetilde{\chi}_{2,3}^0 \to \widetilde{\chi}_{1}^0 a_1$.
It could be much more important than from the decay of the Higgs 
boson or from the associated production.  
In particular, if the gluino and squarks are
light enough for copious production of SUSY events at the LHC, there
would be numerous number of $\widetilde{\chi}_{2,3}^0$ in subsequent
decays of gluinos and squarks.  
The $a_1$ then decays into 
either $b\bar b$ or $\tau^+ \tau^-$ depending on its mass.  Therefore, the
final state will be filled with many $\tau$ leptons, which can be carefully
reconstructed at the mass of $a_1$.  This is a clean, observable, and 
distinguishable signature for NMSSM.
There is also the possibility of $\widetilde{\chi}_{2,3}^0 \to
\widetilde{\chi}_{1}^0 h$ followed by $h \to a_1 a_1$, which again gives 
rise to multi-$\tau$-lepton final states.

\section*{Acknowledgments}
We would like to thank Kok-Wee Song, who was involved in the
initial part of the work.
K.C. thanks the Institute of Theoretical
Physics at the Chinese University of Hong Kong for hospitality.
The work was supported in part by the NSC of Taiwan 
(NSC 96-2628-M-007-002-MY3).


\end{document}